\documentclass[conference]{IEEEtran}
\IEEEoverridecommandlockouts
\usepackage{cite}
\usepackage{amsmath,amssymb,amsfonts}
\usepackage{graphicx}
\usepackage{textcomp}
\usepackage{xcolor}
\usepackage{enumitem}
\usepackage{amsthm}
\usepackage{tikz}
\usepackage{circuitikz}
\usepackage{subcaption}
\usepackage{pgfplots}
\usepackage[capitalise]{cleveref}
\usepgfplotslibrary{groupplots}
\usepackage[binary-units]{siunitx}
\DeclareSIUnit\sample{S}
\pgfplotsset{compat=1.14}

\usepackage{algorithm} 
\usepackage{algpseudocode} 

\usepackage{tikz}
\usetikzlibrary{fit,backgrounds,calc,  automata, positioning, arrows,shapes}
\usetikzlibrary {circuits.logic.US}
\tikzset{circuit logic US}
\definecolor{cbrew0}{RGB}{166,206,227}
\definecolor{cbrew1}{RGB}{31,120,180}
\definecolor{cbrew2}{RGB}{178,223,138}
\definecolor{cbrew3}{RGB}{51,160,44}
\definecolor{cbrew4}{RGB}{251,154,153}
\definecolor{cbrew5}{RGB}{227,26,28}
\definecolor{cbrew6}{RGB}{253,191,111}
\definecolor{cbrew7}{RGB}{255,127,0}
\definecolor{cbrew8}{RGB}{202,178,214}
\definecolor{cbrew9}{RGB}{106,61,154}
\definecolor{cbrew10}{RGB}{20,20,20}

\pgfplotscreateplotcyclelist{allvariantscolorlistevenodd}{%
  {cbrew0, mark = none},  {cbrew0, mark = none,dashed},
  {cbrew1, mark = none},  {cbrew1, mark = none,dashed},
  {cbrew2, mark = none},  {cbrew2, mark = none,dashed},
  {cbrew3, mark = none},  {cbrew3, mark = none,dashed},
  {cbrew4, mark = none},  {cbrew4, mark = none,dashed},
  {cbrew5, mark = none},  {cbrew5, mark = none,dashed},
  {cbrew6, mark = none},  {cbrew6, mark = none,dashed},
  {cbrew7, mark = none},  {cbrew7, mark = none,dashed},
  {cbrew8, mark = none},  {cbrew8, mark = none,dashed},
  {cbrew9, mark = none},  {cbrew9, mark = none,dashed},
  {cbrew10, mark = none},  {cbrew10, mark = none,dashed},
}

\pgfplotscreateplotcyclelist{allvariantscolorlist}{%
  {cbrew0, mark = none}, 
  {cbrew1, mark = none},
  {cbrew2, mark = none}, 
  {cbrew3, mark = none},
  {cbrew4, mark = none},
  {cbrew5, mark = none},
  {cbrew6, mark = none},
  {cbrew7, mark = none},
  {cbrew8, mark = none},
  {cbrew9, mark = none},
  {cbrew10, mark = none},
}

\pgfplotscreateplotcyclelist{navariantscolorlist}{%
  {cbrew1, red!60!black},
  {cbrew3, blue!60!black},
  {cbrew4, green!60!black},
  {cbrew5, orange!60!black},
  {cbrew6, violet!60!black},
  {cbrew7, brown!60!black},
  {cbrew8, cyan!60!black},
  {cbrew9, magenta!60!black},
  {cbrew10, gray!60!black}
}

\pgfplotscreateplotcyclelist{fivevariantscolorlist}{%
  {cbrew1, mark = none},
  {cbrew3, mark = none},
  {cbrew4, mark = none},
  {cbrew8, mark = none},
  {cbrew9, mark = none},
  {cbrew10, mark = none},
}

\usetikzlibrary{shapes,arrows,chains}
\def\BibTeX{{\rm B\kern-.05em{\sc i\kern-.025em b}\kern-.08em
    T\kern-.1667em\lower.7ex\hbox{E}\kern-.125emX}}
\begin{document}

\title{Backing the Wrong Horse: How Bit-Level Netlist Augmentation can Counter Power Side Channel Attacks}

\author{
    \IEEEauthorblockN{Ali Asghar, Andreas Becher, Daniel Ziener}
    \IEEEauthorblockA{Technical University Ilmenau
    \\\{ali.asghar,andreas.becher,daniel.ziener\}@tu-ilmenau.de}}

\maketitle

\begin{abstract}
The dependence of power-consumption on the processed data is a known vulnerability of CMOS circuits, resulting in side channels which can be exploited by power-based side channel attacks (SCAs). These attacks can extract sensitive information, such as secret keys, from the implementation of cryptographic algorithms.
Existing countermeasures against power-based side channel attacks focus on analyzing information leakage at the byte level. However, this approach neglects the impact of individual bits on the overall resistance of a cryptographic implementation. In this work, we present a countermeasure based on single-bit leakage.
The results suggest that the proposed countermeasure cannot be broken by attacks using conventional SCA leakage models.
\end{abstract}


\section{Introduction}
\label{intro}
All physical systems emit observable signals. These signals can also include side channels which (if let unguarded) can provide unintended information pathways for potential attacks. These information channels can be exploited by attackers to extract the sensitive data or implementation details of a system.

Side channels differ from standard communication channels as they are not meant for data transmission and are therefore frequently ignored as potential security risks. A closer examination of attacks exploiting side channels to extract sensitive data reveals a fundamental issue: the core cryptographic principle of \emph{Confidentiality} does not extend to the underlying hardware circuits in the presence of unprotected side channel emissions.

Adversarial attacks have successfully utilized various types of side channels, including sound \cite{sound_counter}, temperature \cite{temp_sca}, light \cite{laser_sca}, timing \cite{kocher-timing}, power consumption \cite{kocher}, and electromagnetic radiation \cite{em_sca}. 

While encryption algorithms may be mathematically secure, their physical implementation on hardware can introduce side-channel vulnerabilities that potentially undermine their overall security.

This study specifically investigates the risks that power-based side channels pose to cryptographic implementations on Field-Programmable Gate Arrays (FPGAs). These attacks take advantage of a key characteristic of Complementary Metal Oxide Semiconductor (CMOS) technology in which  current is drawn only when transistors switch states. This switching activity results in instantaneous power usage, known as leakage, which varies based on the data being processed. As a result, attackers can potentially infer the values by analyzing this power consumption. In the context of this work, leakage refers to the data-dependent voltage fluctuations caused by bit transitions during encryption processes.
As the bits toggle, they consume current, which appears as a voltage drop in the AC-coupled signal of the power supply

Since the introduction of power-based side-channel attacks (SCAs), several countermeasures have been proposed. These countermeasures primarily aim to minimize or eliminate the data-dependent nature of information leakage. Among the widely adopted techniques are masking and hiding. This paper concentrates on hiding-based countermeasures, which work by introducing randomness into the power consumption of cryptographic operations. The goal is to break the correlation between power usage and the underlying data being processed. Hiding is particularly attractive for low-resource devices due to its efficiency in terms of power, performance, and area overhead.

In this work, we present a new hiding-based countermeasure approach which instead of working on byte-level attempts to modify the leakage profile of single bits. Specific bits are patched with a specialized countermeasure circuit designed to disrupt the correlation between power consumption and the underlying data being processed.

The rest of this paper is structured as follows: \Cref{background} provides a summary of the existing hiding-based SCA countermeasures, where \Cref{hiding} explores the basics of \emph{Correlation Power Analysis (CPA)} and examines why current hiding countermeasures fall short in effectively mitigating such attacks. \Cref{proposal} presents the underlying theory of the proposed bit-level countermeasure. The implementation and evaluation of the countermeasure circuit is presented in \Cref{impl}.
Finally, \Cref{conc} concludes the paper and outlines future research directions.

\section{Previous Work}
\label{background}
As discussed in \Cref{intro}, two of the most common methods to enhance resistance against power-based side-channel attacks (SCAs) are \emph{hiding} and \emph{masking}. Masking \cite{sound_counter,kocher_dpa,threshold_nikova,domain_gross} is widely considered one of the most effective defenses against SCAs. Like hiding, masking aims to weaken the relationship between the device’s power consumption and the intermediate values computed during the algorithm's execution. The principle behind masking is to avoid using the actual data values in computations. Instead, these values are obscured through random transformations that can later be reversed to retrieve the original data. This means attackers will capture power traces corresponding to randomized data, making it nearly impossible to link the measurements to the secret data.

In contrast to \emph{Masking}, \emph{Hiding} is a more straightforward countermeasure that focuses on reducing the correlation between the processed data or operations and the resulting power consumption. This is achieved by introducing randomness, either in the time-domain (shifting power peaks over time) or in the power consumption itself (altering the power profile’s structure). Randomness in the time-domain can be introduced by shuffling the order of AES operations, which changes the power profiles across different encryption runs, or by inserting dummy cycles \cite{DPA-book}. Another approach to introduce time-domain randomness is by using an unstable clock source for the cryptographic operations \cite{Moradi}. To alter the power consumption, background noise-generating blocks can be employed, running alongside the encryption process to influence the overall power usage. The goal is to lower the \emph{Signal to Noise (SnR)} ratio by obscuring the signal within the noise. One well-known power-balancing technique in the ASIC domain \cite{power_balance}\cite{logic_dpa}\cite{mask_mdpl} is \emph{Dual-Rail Precharge Logic (DRPL)}. In a DRPL setup, a replica of the encryption algorithm is implemented on the same hardware. The replica operates on complementary data, meaning that when the original circuit processes a data value of '1', its counterpart processes '0'. The idea behind this approach is that, by running both true and complementary data in parallel, the total power consumption will be balanced over time. While implementing a DRPL scheme is relatively straightforward for ASICs, it becomes more challenging for FPGAs \cite{ender_power_equalize,glitch_moradi} due to the heterogeneous nature of FPGA fabrics, which makes it difficult to replicate the encryption circuit perfectly. In \cite{dual_rail_schaumont}, the authors tackle this issue to some degree by suggesting a custom routing algorithm aimed at enhancing the symmetry between the two copies of the design. However, despite these improvements, the resulting implementations still exhibit detectable leakage.

Another category of hiding countermeasures takes advantage of the reconfiguration capabilities of FPGAs, allowing the system to alternate between different physical implementations (variants) of an encryption algorithm \cite{Bosch,nadir,sasdrich_logic_reconf,jlpea2022}. While these variants are functionally equivalent, their structures differ. As the system switches between these variants, an attacker encounters varying power profiles at different points in time, thereby complicating the attack process.

Although hiding countermeasures can mitigate data-dependent leakages, they are only partially effective in enhancing resistance against SCAs and cannot entirely prevent them. If an attacker collects a sufficient number of traces, the countermeasures can still be bypassed. Additionally, basic techniques like averaging or pattern matching \cite{DPA-book} are often enough to undermine the effectiveness of hiding measures.

In the next section, we review the fundamentals of a widely used power-based SCA method called \emph{Correlation Power Analysis (CPA)} to better understand the inherent limitations of hiding countermeasures.

\section{Limitations of Hiding against CPA}
\label{hiding}
As mentioned earlier, power attacks leverage the relationship between the device's instantaneous power consumption and an intermediate value (processed data) generated by the cryptographic algorithm at a specific moment in time.

\emph{Correlation Power Analysis (CPA)} is a widely used SCA technique that seeks to identify the correlation between power consumption and intermediate values by estimating the device's hypothetical power usage through a leakage model. This model, which is based on the data and key, predicts the power consumed during an intermediate operation. The predicted values are then correlated to the actual power traces measured from the device during the execution of the cryptographic algorithm to find correlations. For a power attack to succeed, a precise model is essential to estimate the power consumption during intermediate operations, which requires insight into the internal workings of the algorithm and its specific implementation on the device. According to \cite{DPA-book}, the \emph{Hamming Distance (HD)} metric is considered an effective power model for targeting hardware-based cryptographic systems, as it better reflects the complexities of CMOS circuit power usage. A well-known HD-based attack involves measuring the state difference between the 9th and 10th rounds of AES-128. This is achieved by identifying the bits that change in the state register during the final round of AES encryption. To find these transitioning bits, the following equation can be used:

\begin{equation} \label{eq:toggle-count}
    \begin{split}
      {Transitions}= & Sbox^{-1}[SR^{-1}[CT [i][j] \oplus 
                          \\ & Key_{LR}[j]]] \oplus CT [i][j]
    \end{split}
\end{equation}

where, $CT$, and $Key_{LR}$ are the abbreviations for \emph{Cipher Text} and \emph{Last Round Key}, respectively. $Sbox^{-1}$ and $SR^{-1}$ represent the inverse switch-box and shift-rows operations respectively. \emph{CT[i][j]} represents the $j^{th}$ cipher text byte corresponding to $i^{th}$ encryption.

Now, to calculate the HD, we take the Hamming weight (HW) of the transitions from \Cref{eq:toggle-count}.

The estimated power consumption based on the HD value of the intermediate result calculated using \Cref{eq:toggle-count} is calculated as follows:
\begin{equation} \label{eq:model}
    \begin{split}
      HD= & HW(Transitions)  \\ 
      {Hypothetical\_Power} = & HD 
    \end{split}
\end{equation}

It is important to note that this attack model operates under the assumption that each bit toggle contributes \textbf{equally} to the overall power consumption.
To simplify the attack, the model focuses on one key byte \emph{j} at a time. The process is then repeated for remaining bytes to fully recover the entire key.

For hardware implementations, data-dependent leakage signals are already suppressed by a considerable amount of background switching noise. Yet, the CPA excels in extracting the secret key from the measured power traces. Therefore, hiding, which adds noise to encryption operations simply increases the attack complexity (by decreasing SNR) which is manifested by the increased number of traces required to obtain the secret key. The effectiveness of CPA stems from its ability to sift through randomness by evaluating potential key-byte values, from a range of 0 to 255, and ranking them according to their correlation values. The key-byte values that show stronger correlations are assigned higher probabilities. This process is iteratively refined using the collected power traces, progressively narrowing down to a high-confidence estimate of the correct key-byte.

The results from the studies (related to hiding) referenced in \Cref{background} indicate that hiding countermeasures improve resistance to CPA attacks, as they require a larger number of traces compared to an unprotected AES implementation before the eventual recovery of the correct key.

To conclude, while hiding offers some improvement in security against attacks like CPA, its effectiveness is limited. Since hiding works by reducing the Signal-to-Noise Ratio (SnR) to complicate correlation attacks, it is practically impossible to entirely obscure the signal within the noise to achieve an SnR $\sim 0$. 

\section{Proposed Countermeasure}
\label{proposal}
In the previous section, we discussed the limitations of using hiding as a defense against CPA. The effectiveness of CPA lies in its ability to differentiate the correct key from incorrect ones, even when there is noise present. For our proposed countermeasure, rather than concealing the correct key-byte under noise, we add extra leakage (which we refer to as an \textbf{offset}) that causes incorrect key values to appear as strong contenders for the best guess. 
To achieve this, it is necessary to disrupt the correlation between the model's estimated leakage (HD values) and the actual leakage observed from the device. As noted earlier, the correlation with HD values is based on the assumption of a fixed offset for each bit transition.
This correlation can be disrupted by inverting the leakage behavior of a specific bit. Instead of introducing higher leakage when a particular bit toggles, we deliberately add an artificial offset when that bit does not toggle. The offset must be large enough to distort the typical Hamming Distance (HD) pattern, for example, if one specific bit (e.g., bit '1') along with three others toggles (resulting in HD=4), the added offset should reduce the observed leakage to a level consistent with only one or fewer bit transitions (HD$\leq$1).

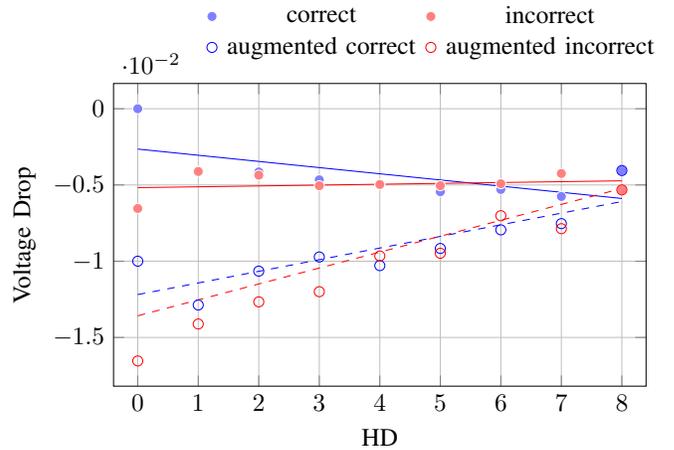
\begin{figure}
\begin{tikzpicture}[scale=0.95] 
    \begin{axis}[
        xlabel={HD},
        ylabel={Voltage Drop},
        legend style={at={(0.6,1.05)},fill=none,anchor=south,draw=none,column sep=0.2em},
        legend columns=2,
        enlarge x limits=0.05,
        xtick={0,1,...,8},
        grid=major,
        height=0.8*\axisdefaultheight,
        width=1.02*\columnwidth,
        xmin = 0,
        xmax=8,
    ]
        \def\o{0.07062413832720588} 
        \addplot [draw=white,fill=blue!50,only marks,mark=*] table [col sep=comma, x={hd}, y expr={\thisrow{corr}+\o}] {data/v29_bit2_off0_key180.txt};
        \addlegendentryexpanded{correct};
        
        \addplot [draw=white,fill=red!50,only marks,mark=*] table [col sep=comma, x={hd}, y expr={\thisrow{best}+\o}] {data/v29_bit2_off0_key33.txt};
        \addlegendentryexpanded{incorrect};
        
        \addplot [blue,only marks,mark=o] table [col sep=comma, x={hd}, y expr={\thisrow{corr}+\o}] {data/v29_bit2_off10_key180.txt};
        \addlegendentryexpanded{augmented correct};
        
        \addplot [red,only marks,mark=o] table [col sep=comma, x={hd}, y expr={\thisrow{best}+\o}] {data/v29_bit2_off10_key33.txt};
        \addlegendentryexpanded{augmented incorrect};

        \addplot  [mark=None, color=blue, domain=0:8,] {-0.0004048838393120574*x+ -0.002646623089935382};
        \addplot  [mark=None, color=blue,dashed, domain=0:8,] {0.0007629070187212877*x- 0.012188916303942693};
        
        \addplot + [mark=None, color=red,solid, domain=0:8,] {5.604989262412522e-05*x -0.005176080833122899};        
        \addplot + [mark=None, color=red,dashed, domain=0:8,] {0.0010438902088659857*x -0.01358071697358815};
        
    \end{axis}
\end{tikzpicture}
\caption{
Measured voltage drop of the power supply for byte 0 over the HD values.
\textcolor{blue}{Blue} are the values for the correct key while \textcolor{red}{Red} are the values for an incorrect key.
Lines indicate the resulting fitted lines of the HD points.
Dashed versions indicate the result of an augmented implementation. 
One can see that in the augmented case an incorrect key would yield a better correlation}
\label{hd-leakage}
\end{figure}

The concept is further illustrated in \Cref{hd-leakage}, which demonstrates the relationship between leakage values and the Hamming Distance (HD) of byte 0 during the state register overwrite operation between the 9th and 10th rounds of AES. In this context, \emph{augmented} refers to a hardware implementation that introduces additional leakage as described earlier. For the correct key guess (represented by the solid blue line), various HD values (indicated by the filled blue circles) align along a straight line with a negative slope, suggesting that power consumption increases as the HD grows. This relationship results in a high correlation coefficient \emph{r $\sim 0.645$}.
In contrast, the plot with the solid red line, representing an incorrect key guess, shows that the HD values do not align well with the leakage, and the slope is close to zero, indicating no change in leakage as the HD increases. Next, we examine the effect of introducing an offset to the leakage of a single bit (bit 2, when it does not toggle), represented by the dashed lines fitting the unfilled data points. A noticeable shift in the slope to a positive value occurs, signifying a decrease in voltage drop as the HD increases, which is consistent with the explanation provided earlier. Interestingly, the key-byte that was previously deemed incorrect (solid red line) now exhibits much stronger linearity and a positive slope (dashed red line), with a correlation coefficient of \emph{r $\sim 0.979$}. This is higher than the correlation (\emph{r $\sim0.848$}) for the correct key with an offset (dashed blue). 
As a result, the CPA attack is no longer able to successfully recover the correct key.

It is important to highlight that, unlike hiding, which introduces random noise into the power leakage, the offset is a fixed value applied only when a specific bit in a state byte does not toggle. This approach is crucial because adding random noise to disrupt the HD alignment would ultimately be filtered out by the CPA, given its ability to handle and bypass randomness, as discussed in \Cref{hiding}.

In the following section, we share the implementation details and evaluated results of our offset generation countermeasure circuit which can augment specific bits (when they toggle) with an offset value to disturb the HD alignment.

\begin{figure*}
    \begin{subfigure}[c]{0.49\textwidth}
    \begin{tikzpicture}[scale=0.95]
        \begin{axis}[
                width=\columnwidth-0.5em,
                height=\axisdefaultheight,
                ylabel={Correlation},
                xlabel={Number of traces},
                grid=major,
                ymin=0.01,ymax=0.4,
                    enlargelimits=false
                ]
            \foreach \c in {0,...,254} {
                    \addplot[forget plot,gray] table [col sep=comma,x={key},y={cpa_value}] {data/key\c_cpa_vals.txt};
            }
            \addplot[gray] table [col sep=comma,x={key},y={cpa_value}] {data/key255_cpa_vals.txt};
            \addlegendentry{other};
            \addplot[blue,line width=1.25pt] table [col sep=comma,x={key},y={cpa_value}] {data/key180_cpa_vals.txt};
            \addlegendentry{correct};
            
            \end{axis}
        \end{tikzpicture}
        \caption{without countermeasure}
        \label{fig:ohne-counter}
    \end{subfigure}
    \begin{subfigure}[c]{0.49\textwidth}
    \begin{tikzpicture}[scale=0.95]
        \begin{axis}[
                width=\columnwidth-0.5em,
                height=\axisdefaultheight,
                ylabel={Correlation},
                xlabel={Number of traces},
                grid=major,
                ymin=0.01,ymax=0.4,
                    enlargelimits=false
                ]
            \foreach \c in {0,...,254} {
                    \addplot[forget plot,gray] table [col sep=comma,x={key},y={cpa_value}] {data/key\c_cpa_vals_v129210705.txt};
            }
                    \addplot[gray] table [col sep=comma,x={key},y={cpa_value}] {data/key255_cpa_vals_v129210705.txt};
            \addlegendentry{other};
            \addplot[blue,line width=1.25pt] table [col sep=comma,x={key},y={cpa_value}] {data/key180_cpa_vals_v129210705.txt};
            \addlegendentry{correct};
            
            \end{axis}
        \end{tikzpicture}
        \caption{with countermeasure}
        \label{fig:mit-counter}
    \end{subfigure}
        \caption{Correlation of for all possible values of key byte 0 with an increasing number of traces with and without countermeasure}
\end{figure*}
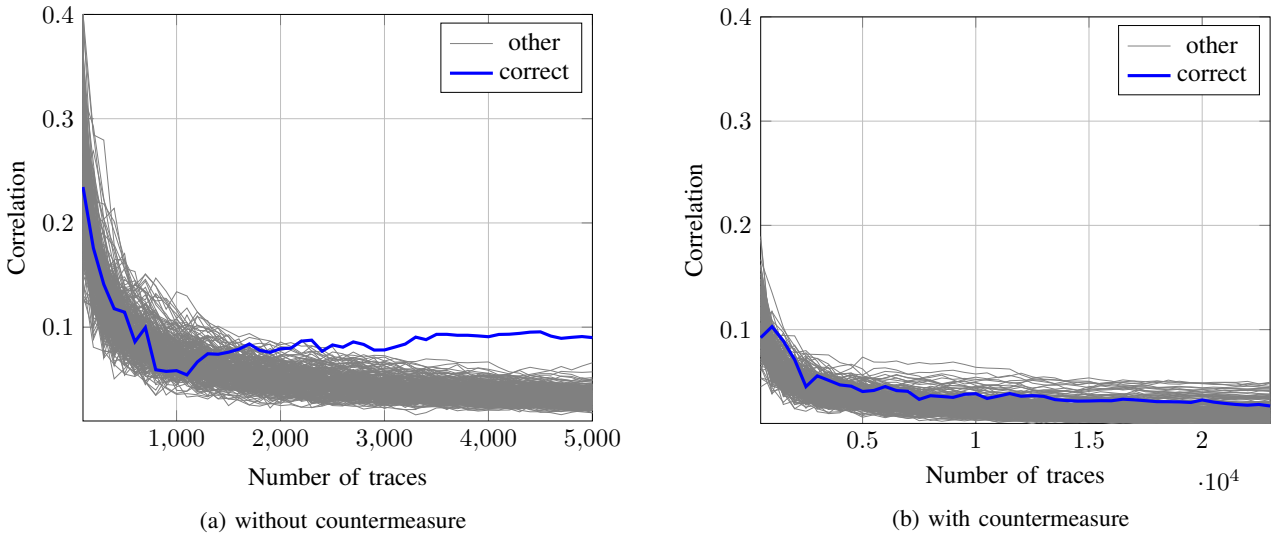

\section{Implementation and Evaluation}
\label{impl}

In this section, we provide a brief overview of the hardware implementation of the proposed offset generation circuit, followed by the details of the measurement setup and CPA results to evaluate the SCA resistance of the proposed countermeasure. 

For the countermeasure circuit, we chose a bank of \emph{Ring Oscillators (ROs)} driven by a tapped delay line. 
This design was selected due to its flexibility in adjusting the amount of the applied offset as well as the duration the offset is applied. 
The amount is controlled by the number of implemented and enabled ROs (\emph{\#RO}) while the timespan the ROs are active can be modified by altering the effective length of the delay line \emph{L}.
A XNOR LUT selects two of the 6 inputs provided from the taps of the delay line to generate the pulse that enabled the ROs. 
The inputs of the LUT selected for the XNOR operation can be changed by reconfiguring the INIT value of the LUT.

While more efficient designs for generating the required offset values may be possible, our focus in this work was on demonstrating the proof-of-concept and validating the functionality of the countermeasure at the hardware level. Optimizing the offset generation circuits will be a key area for future work.

\subsection{Experimental Setup}
\label{traces}
To evaluate the resistance of our proposed countermeasure circuit, we utilize the ChipWhisperer (CW) \cite{CW-white} measurement setup.

For this work, we use the CW305 target board which features an Artix-7 (xc7a100t) target device and has been interfaced with ChipWhisperer-Lite (capture board) for collecting power traces and carrying out SCA. For more information on the CW305 and CW-lite boards, please refer to \cite{CW-white}.
Further details of the measurement setup are as follow:
\begin{enumerate}
    \item The AES core used for experimentation was obtained from the ChipWhisperer repository \cite{cw_git}, which is a simple realization of an AES-128.
    \item The operating frequency of the AES implementations is \SI{10}{\mega\hertz}.
    \item The power traces for encryption have been collected by measuring the voltage-drop (amplified by 20 dB) in the ac-coupled signal of the power supply.
    \item A single encryption run corresponds to one measured trace.
    \item The traces are passed to the capture board via a SMA connector and are then sampled at \SI{40}{\mega\sample\per\second}.

\end{enumerate}

\subsection{CPA Results}
To assess the effectiveness of our countermeasure against CPA, we focus on attacking the last round state register overwrite operation of AES. The leakage values associated with this operation were estimated using the HD-based model, as described in \Cref{eq:toggle-count}.

The results shown in \Cref{fig:ohne-counter} and \Cref{fig:mit-counter} illustrate the outcomes of the CPA attack on AES implementations both with and without the proposed countermeasure. From the data in \Cref{fig:ohne-counter}, it is evident that without any countermeasure, the correct key byte0 value (180) can be identified with approximately 3000 traces. In contrast, when the AES implementation is modified by patching bit 2 of byte 0 with the proposed countermeasure characterized by \emph{\#RO=70} and utilizing input \emph{I5} of the XNOR LUT, the resistance to the attack improves significantly. The use of the XNOR gate in the countermeasure circuit is designed to minimize the power consumption associated with toggling of the augmented bit (see \Cref{proposal}). Since the XNOR gate outputs 0 when its inputs are equal, if the augmented bit toggles, the output remains at 0 for a duration determined by distance between the delayline taps used by the XNOR LUT.
The measurement results shown in\Cref{fig:mit-counter} indicate that the correct key byte value was not recovered for $\sim 23000$ traces. We also tested our countermeasure against CPA with an increased number of traces, successful key recovery wasn't possible for $\sim 46000$.

\section{Conclusion \& Future work}
\label{conc}
Traditional countermeasures for hiding, which focus on shielding the correct key by introducing noise into the power consumption during cryptographic operations, offer only limited protection against power-based side-channel attacks (SCAs). In this paper, we propose a novel approach for securing cryptographic implementations against power-based SCAs. Our method involves intentionally exposing incorrect key values as potential candidates for the best key guess in a CPA attack by adding an offset to a specific bit when it remains static. 

In the future, we aim to investigate alternative circuit designs for our proposed countermeasure that offer improved area and energy efficiency.

\bibliographystyle{ieeetr}
\bibliography{arxiv_paper.bib}

\end{document}